\documentclass[12pt,apj]{emulateapj}
\shorttitle{Resolving Altair}
\shortauthors{Peterson \etal}
\slugcomment{Submitted to the {\em Astrophysical Journal} June 9, 2005\\
             Revised draft of \today}
\usepackage{graphics}
\newcommand{\ie}{i.e.,}
\newcommand{\eg}{e.g.,}
\newcommand{\etal}{et~al.}

\newcommand{\um}{\ensuremath{\mu\mbox{m}}}
\newcommand{\kms}{\ensuremath{\mbox{km\,s}^{-1}}}

\newcommand{\Msun}{\ensuremath{\mbox{$M_{\sun}$}}}
\newcommand{\Rsun}{\ensuremath{\mbox{$R_{\sun}$}}}
\newcommand{\Lsun}{\ensuremath{\mbox{$L_{\sun}$}}}
\newcommand{\decasec}{\ensuremath{\,.\!\!''}}

\newcommand{\ddeg}{\ensuremath{.\!\!^{\circ}}}
\newcommand{\dhrs}{\ensuremath{.\!\!^h}}
\newcommand{\bea}{\begin{eqnarray*}}
\newcommand{\ba}{\begin{array}}
\newcommand{\bdm}{\begin{displaymath}}
\newcommand{\eea}{\end{eqnarray*}}
\newcommand{\ea}{\end{array}}
\newcommand{\edm}{\end{displaymath}}
\received{2005 June 9}
\begin{document}

%\input{TitlePage}
%\vfill\eject
\title{Resolving the Effects of Rotation in Altair \\
    with Long-Baseline Interferometry}

\author{D.~M.\ Peterson\altaffilmark{1}, C.~A.\ Hummel\altaffilmark{2,4},
T.~A.\ Pauls\altaffilmark{3}, J.~T.\ Armstrong\altaffilmark{3},
J.~A.\ Benson\altaffilmark{5}, C.~G.\ Gilbreath\altaffilmark{3},
R.~B.\ Hindsley\altaffilmark{3}, D.~J.\ Hutter\altaffilmark{5}, 
K.~J.\ Johnston\altaffilmark{4}, D.\ Mozurkewich\altaffilmark{6},
\& H.\ Schmitt\altaffilmark{3,7}}

\altaffiltext{1}{Department of Physics and Astronomy, Stony Brook University, Stony Brook, NY 11794-3800 email: dpeterson@astro.sunysb.edu}
\altaffiltext{2}{European Southern Observatory (ESO), Casilla 19001, Santiago 19, Chile email: chummel@eso.org}
\altaffiltext{3}{Naval Research Laboratory, Code 7215, 4555 Overlook Ave.~SW, Washington, DC 20375 email: pauls@nrl.navy.mil, tom.armstrong@nrl.navy.mil, hindsley@nrl.navy.mil, henrique.schmitt@nrl.navy.mil}
\altaffiltext{4}{U.S.\ Naval Observatory, 3450 Massachusetts Ave.~NW, Washington, DC, 20392-5420 email: kjj@astro.usno.navy.mil}
\altaffiltext{5}{U.S.\ Naval Observatory, Flagstaff Station, 10391 W.~Naval Observatory Rd., Flagstaff, AZ 86001-8521 email: jbenson@nofs.navy.mil, djh@nofs.navy.mil}
\altaffiltext{6}{Seabrook Engineering, 9310 Dubarry Rd., Seabrook, MD 20706 email: dave@mozurkewich.com}
\altaffiltext{7}{Interferometrics, Inc., 13454 Sunrise Valley Drive, Suite 240, Herndon, VA20171}
\begin{abstract}
placeholder
\end{abstract}

%\input{Abstract}
%\vfill\eject

\begin{abstract} 

We report successful fitting of a Roche model, with a surface
temperature gradient following the von Zeipel gravity darkening law, to
observations of Altair made with the Navy Prototype Optical
Interferometer.  We confirm the claim by Ohishi, Nordgren, \& Hutter
that Altair displays an asymmetric intensity distribution due to
rotation, the first such detection in an isolated star.  Instrumental
effects due to the high visible flux of this first magnitude star
appear to be the limiting factor in the accuracy of this fit, which
nevertheless indicates that Altair is rotating at $0.90\pm0.02$ of its
breakup (angular) velocity.  Our results are consistent with the
apparent oblateness found by van Belle \etal\ and show that the true
oblateness is significantly larger owing to an inclination of the
rotational axis of $\sim 64^{\circ}$ to the line of sight.  Of
particular interest, we conclude that instead of being substantially
evolved as indicated by its classification, A7\,VI-V, Altair is only
barely off the ZAMS and represents a good example of the difficulties
rotation can introduce in the interpretation of this part of the HR
diagram.

\end{abstract}


%\input{Intro}
%\vfill\eject

\section{INTRODUCTION}\label{Introduction}

Altair (variously $\alpha$\,Aql, 53\,Aql, HR\,7557, HD\,187642, of
spectral type A7\,VI-V) is one of the brightest stars in the Northern
sky, sharing membership in the ``Summer Triangle'' with two other
notable A stars.  Unlike Vega and Deneb, Altair shows a rather diffuse
spectrum which was early recognized to be due to a large projected
rotational velocity variously estimated at 242\,\kms\
\citep{UesugiFukuda1982}, 217\,\kms\ \citep{Royer02}, and 200\,\kms\
\citep{AbtMorrel95}.  These estimates of its projected velocity, a
lower limit for the true rotational velocity, are already a significant
fraction of the breakup velocity, estimated near 400\,\kms.

Altair has become a significant object in understanding the atmospheres
of main sequence stars at masses near but above that of the Sun.
Specifically, Altair and $\alpha$\,Cep are the two hottest stars
showing Ly$\alpha$\ and \ion{C}{2} emission, taken as indicators of a
chromosphere \citep{Simon94, Walter95}.  The absence of these
indicators at earlier spectral types is taken to mean that significant
convection disappears at this point on the upper main sequence.  We
note that $\alpha$\,Cep also has a high projected rotation velocity
with estimates of 246\,\kms\ \citep{UesugiFukuda1982}, 196\,\kms\
\citep{Royer02}, and 180\,\kms\ \citep{AbtMorrel95} listed.

Altair's known high rotation rate has prompted attempts to measure the
geometrical effects of its rotation over the years, starting with the
Intensity Interferometer \citep{Narrabri74}.  However, it was not until
the near-IR observations with the Palomar Testbed Interferometer
\citep[PTI;][]{PTI} by \citet{vBelle01} that a significant flattening
was detected.  Comparison to classical \citep{vZ24} Roche models showed
the flattening was completely consistent with the observed projected
rotation.

Although this agreement between theory and observation is nothing short
of epochal, it is incomplete.  Except near breakup, apparent 
oblateness, interpreted through Roche theory, displays the same
degeneracy between equatorial velocity and tilt (inclination) as the
apparent rotation velocity: one determines the quantity $v_{eq}\sin i$
well, but not the two separately.  Nor can one determine the sense of
rotation, pro- or retrograde.  Besides providing a test of the
flattening predicted by theory, oblateness measurements do yield the
position angle of the angular momentum vector, the projection of that
vector on the plane of the sky.

In addition to flattening, \citet{vZ24} predicted that for moderate
rotation stellar disks would display variable surface temperatures,
hotter on the rotational axes and cooler at the equator.  Specifically,
if one defines a local effective gravity accounting for centrifugal
acceleration, then the local effective temperature is related to the
effective gravity as $T_{eff}^4 \propto g_{eff}$, which is referred to
as ``gravity darkening''.  With sufficient rotation and at intermediate
inclinations, gravity darkening predicts that stellar disks will
display asymmetric intensity distributions.

As we describe below, this prediction is of great interest in the field
of optical interferometry.  Asymmetric intensity distributions produce
significant imaginary components in the visibilities, usually
represented as a non-trivial visibility phase.  Recently developed
techniques for recovering a closely related quantity, ``closure phase''
\citep{COAST,NPOI-Close}, are now being applied to the first round of
stellar objects, \citep[\eg][]{Witt01}.

Although originally proposed as a follow-up on the oblateness
observations, Altair was observed at the Navy Prototype Optical
Interferometer \citep[hereafter NPOI,][]{NPOI} while the three beam
combiner was in operation, allowing measurement of closure phase around
one complete triangle.  Examination of the data immediately revealed
the intermediate phase angles, unambigiously signaling the presence of
an asymmetric intensity profile \citep{Ohishi03}.

Using a model consisting of a limb-darkened disk and a bright spot,
\citet{Ohishi04} demonstrated both the previously discovered oblateness
and the necessity of including asymmetries in the intensity
distribution.  They argued that the probable interpretation was that of
rotational flattening and gravity darkening.

In the meantime we have become aware of some limitations in those data
due to inadequate corrections for ``deadtimes'' in the avalanche
photodiode detectors which affect the high signal levels from objects
as bright as Altair.  We therefore reconsider a subset of these data
that is relatively immune to the detector problems, using a full
implementation of von Zeipel's theory \citep{vZ24} for the model
fitting, and redoing the reductions in a way that dramatically reduces
noise in the bluest channels.  We find that a Roche model rotating at
90\% of the breakup angular velocity and inclined $\sim64^{\circ}$ from
pole-on fits the observations with high fidelity.

We show that the parameter that sets the overall temperature scaling
for the model, the effective temperature at the poles, $T_p$, is close
to 8700\,K for this model, and the polar surface gravity is
correspondingly fairly high.  This suggests that Altair is less evolved
than one might naively expect from its spectral type and luminosity
classifications.

In this model the equator is 1850\,K cooler than the pole.  Given that
the model includes both polar brightening and a long equatorial swath
of low intensity, this is a complex intensity distribution, and the
agreement with the observations is a strong endorsement for the simple
\citet{vZ24} theory.

Below, we describe the new reductions, give a brief review of Roche
theory and then present the fits.  We note that the existence of large
amounts of surface at near-solar temperatures suggests that the role of
Altair (and probably $\alpha$\,Cep) in defining the high temperature
end of convection on the main sequence may need to be reconsidered.  We
also note the recent announcement that Altair is a low amplitude
($\delta$~Sct) pulsating star \citep{Buzasi05}, which may give hope
that asteroseismology will be able to put useful limits on any gradient
of the angular velocity in the outer envelope.

%\input{Observations}
%\vfill\eject

\section{OBSERVATIONS}

Altair was observed on four nights, 25--27 May and 1~June of 2001,
with the NPOI.  These are the same observations used by
\citet{Ohishi03} and \citet{Ohishi04}; we refer the reader to those
papers for a journal of observations and a description of the observing
details, but we briefly reprise them here.  We have focused here on the
data set obtained May 25, 2001.  This is by far the largest set of
data, while the other data do not increase the range of hour angles
observed in the first night.

The observations used the Astrometric West (AW), Astrometric East
(AE), and West~7 (W7) stations, forming a triangle of interferometric
baselines with lengths of 37.5\,m (AW--AE), 29.5\,m (W7--AW), and
64.4\,m (AE--W7).  The backend combined these three input beams to
produce three output beams, with one baseline on each.  The output
beams were dispersed into 32 spectral channels covering
$\lambda\lambda 443-852$~nm, although the bluest four channels
($\lambda\lambda 443-460$~nm) of the W7--AW output were not
functioning.

The Altair observations were interleaved with observations of a
calibrator, $\zeta$~Aql (A0\,V), about 12$^\circ$ away on the sky.  We
initially estimated its diameter to be around 0.85\,mas, with which it
would have acted as a quite acceptable calibrator.  However, as noted
by \citet{Ohishi04}, $\zeta$~Aql is a rapid rotator with values of
345\,\kms\ \citep{UesugiFukuda1982} and 317\,\kms\ \citep{Royer02}
reported.  We adopted 325\,\kms, raising the question of aspect
dependent corrections to the squared visibilities and phases. This
possibility was discussed by \citet{Ohishi04} who concluded that the
effects were not important at the level of the analysis they conducted,
but who cautioned that the problem needed to be reconsidered if a
detailed analysis of Altair was attempted with these observations.

We have found a number of occasions in 2004 when $\zeta$~Aql was
observed with a second calibrator, $\gamma$~Lyr (B9\,III). $\gamma$~Lyr
is a relatively slow rotator \citep[$\sim70$\,\kms,][]{Royer02} and a bit
fainter ($V\sim3.24$) which, coupled with slightly higher temperature
leads to the expectation of a symmetric, nearly unresolved calibrator
for $\zeta$~Aql.  This turned out to be correct, and to our surprise,
we found that we were able to deduce a meaningful fit of a Roche model
to $\zeta$~Aql using the phase and visibility amplitude data.  Since
such results are rare, our report on Altair here is the first, we have
decided to present a detailed discussion of the case in a separate
communication \citep{P-ZAql}.  

We summarize in columns 6 and 7 of Table \ref{Tab:Models} a preliminary
set of the relevant Roche parameters for $\zeta$~Aql.  Given the small
angular diameter, the quoted errors produce uncertainties in the
calibration of Altair that are undetectable compared to other error
sources.  One property of note from these parameters is the near
equator-on orientation of the spheroid.  At most we would have expected
a few degrees uncertainty in the triple phases induced by the range of
possible inclinations.  However, the observations calibrated by
$\gamma$~Lyr show that these phases are truly small, indistinguishable
from zero at the $\pm1^{\circ}$ level.  The star is apparently seen at
nearly $i\sim90^{\circ}$.

\citet{Ohishi04} used contemporaneous observations of Vega as a check
star - one that should show a circular outline.  Although we agree
those observations do seem to imply that Vega is circular, we are not
inclined to place much weight on the result.  Vega is twice as bright
as Altair in this wavelength range.  Detector nonlinearities, already
serious in the Altair observations as we describe in \S
\ref{Datareduction}, are overwhelming here.  We simply do not know how
to interpret the observed squared visibilities.

Finally, we note that since the NPOI records phases and visibilities
over a range of wavelengths, there is little possibility of
contamination by unknown companions, at least those within the
~0\decasec5 field of view of the siderostats.  If such existed there
would be clear, strongly modulated phases and visibilities.
$\zeta$~Aql is a known member of a wide multiple system, ADS12026, with
companions no closer than 5 arcseconds and magnitude differences no
less than 8.

%\input{DataReduction}
%\vfill\eject

\section{DATA REDUCTION}\label{Datareduction}

\subsection{Incoherent Integration}

The NPOI observes interference fringes by modulating the optical path
on the delay line for each array element, using a triangle-wave pattern
at a frequency of 500\,Hz.  The resulting modulation of the intensity
is detected in 8 bins evenly spaced over one fringe in each channel by
Avalanche Photo Diodes (APD). The phase of the intensity modulation
changes on time scales of milliseconds since the fringe tracker
employed by NPOI tracks the envelope of the (bandwidth limited) fringe
packet rather than the fringe phase.

The data in the delay bins were processed to produce the
complex visibility and squared visibility modulus $V^2$ for each
baseline at each wavelength.  From these the triple product $V_{123}
\exp{i\phi_{\rm cl}}$ can be calculated, where the triple amplitude
$V_{123}=|V_1||V_2||V_3|$ is the product of the amplitudes of the
complex visibilities of the individual baselines and the closure phase
$\phi_{\rm cl}=\phi_1+\phi_2+\phi_3$ is the sum of the individual
phases.  Although the baseline phases themselves are affected by
atmospheric turbulence, those effects cancel in the sum of three phases
around a closed triangle, so the closure phase preserves information
about the source structure.  These data products are produced for each
2\,ms cycle of delay modulation.  In the standard incoherent
integration as described by \citet{Hummel98}, the squared visibilities
and complex triple products are summed to provide average values in one
second intervals.

\subsection{Coherent Integration}

We employed a new algorithm for the coherent integration of the complex
visibilities of the NPOI first presented by \citet{Hummel03}.  Compared
to the incoherent integration of the squared visiblities, coherent
integration achieves a higher SNR of the averages due to the larger
number of photons detected in a coherent sample of the fringe. We have
exploited this fact to recover meaningful results from all NPOI
spectrometer channels, while the channels on the blue side of about
560\,nm had usually been discarded in incoherent reductions due to the
insufficient number of photons detected during a 2\,ms instrumental
integration time. For the coherent integration time we selected
200\,ms, and the resulting complex visibilities were both combined to
form complex triple products and transformed individually into squared
amplitudes of the modulus. Every ten samples of these quantities were
then averaged (averaging real and imaginary part of the complex triple
products separately) for a total integration time per data point of
2\,s.

The alignment of the raw visibility phasors neccessary before integration
in order to avoid detrimental coherence losses was performed as follows.
Two steps are necessary to rotate the phasors onto a common fringe in
order to enable a phase tracking algorithm.

First, average power spectra of the channeled visibility as a function 
of delay were computed for 10 ms intervals. Their maxima, corresponding
to the group delays of the fringe packets,
are not zero but have a typical RMS on the order
of one micron as the NPOI group delay fringe
tracker tries to center the fringe, but does not lock onto its phase.

Second, from the deviation in position of the fringe from
the estimated geometrical value, which is non-zero and has a typical RMS
on the order of 10\,\um{} due to atmospheric refractive index
fluctuations, we estimated the differential amount of air and thus
the phase shift between the peak of the envelope
of the fringe packet and the nearest fringe peak.
In other words, this phase is the phase of the complex Fourier transform
of the visibility as a function of wavenumber. The modulus of this
transform peaks at the value of the group delay. The phase of the transform
at this delay is called the group delay phase. 

We converted the group delay phase to a delay using the mean wavelength
of the white-light fringe, and added it to the group delay. Rotation of
the visibility phasors of different channels by an angle
corresponding to the ratio of this delay value and the 
wavelength of the spectrometer channel will
align them on the same fringe. At this point, the algorithm
implements a photon-noise limited
off-line fringe phase tracker enabling the use of much longer coherent
integration times.

\subsection{Baseline Bootstrapping}

We used an important modification of the above procedure by applying the
baseline bootstrapping method, a design feature of the NPOI interferometer
\citep{NPOI}. It exploits the fact that the sum of the fringe delays
along a closed loop of baselines is zero (if the same fringe is
identified on each baseline). Therefore, if a long baseline in a
multi-telescope array sees a low contrast fringe due to, e.g., object
extension, and this baseline involves two telescopes which are at the
same time involved with other telescopes of the array on much shorter
baselines seeing much higher fringe contrast, the fringe delay of the
long baseline can be computed from the fringe delays on the shorter
baselines which are ``bootstrapping'' the long one. In the simple case
of the observations described on Altair, the fringe delay on the long
64\,m W7--AE baseline is just the difference between the fringe delays
on the shorter AE--AW and W7--AW baselines.

\subsection{Averaging and Editing}

The 2\,s data points are edited for outliers as described by
\citet{Hummel98}.  The final averaging is done over the full length of
a pointing (called a scan at NPOI) which lasts typically 90\,s. The
computation of the formal errors also follows \citet{Hummel98}, except
that we have implemented a different approach for the complex triple
products based on a suggestion by D.\ Buscher (priv.\ comm.). Under
simple assumptions, the error of a complex triple product is described
by an error ellipse which has one axis aligned with the triple product
phasor. Assuming this, we compute the error of the triple product as the
error of the imaginary and real parts of the mean after applying a
rotation of all phasors by the mean triple product phasor. The errors
of amplitude and phase of the triple product are then equal to the
error of the real part and the error of the imaginary part divided by
the respective amplitudes.

\subsection{Detector Non-linearity}

A source of systematic error comes from deadtime in the pulse counting
electronics controlling the avalanche photodiode detectors.  These
systems saturate at about 1\,MHz and display significant non-linearity
in the apparent count rates as they near this limit.  The nominal
design of the detector systems included a $\tau=200$\,ns deadtime, but
we have subsequently found that not only do those time constants vary
significantly, channel to channel, but also that in a given channel
they depend on the mean signal level because of the effects of heating.
We believe that it will be possible to model and remove these effects,
but some effort is involved, which we will report on in the future.
Unfortunately, these problems, which do not affect the fainter objects
usually observed by the instrument, were not recognized at the time the
Altair observations were made.

However, we believe that through a rather unique set of circumstances
the phase and some of the amplitude measurements acquired during the
2001 observations are to first order free of the effects of these
non-linearities.  One reason was that during these observations only
three stations were in use and the three spectrographs recorded single
baseline data.

The other reason was that the amplitude and phase measurements from
each channel were accomplished with a simple Discrete Fourier
Transform.  By dithering the optical delay at a frequency of $\omega$,
the signal was modulated according to
\begin{equation}
I(t) = I_0 \left[1 + V \cos \left(\omega t + \phi \right) \right]
\label{eq:dither}
\end{equation}
where V is the (instrumental) amplitude of the visibility and $\phi$
the instantaneous phase.  The detector system responded to the
modulated signal according to
\begin{equation}
N(t) = \frac{Q I(t)}{1+\tau Q I(t)} \sim N_0(t)-\tau N_0^2(t)+\cdots.
\label{eq:deadtime}
\end{equation}
Where $Q$ is the quantuum efficiency, $N(t)$ is the apparent pulse rate
and $N_0(t)= Q I(t)$ is the true photon detection rate.  In the
linearized form we assumed $\tau N_0 << 1$, which for a 300\,kHz count
rate, typical of the wider red channels on the higher visibility
baselines, is adequate to 1\% or better.  Substituting equation
\ref{eq:dither} into \ref{eq:deadtime}, clearing the quadratic cosine
using the half angle formula and collecting terms, this becomes
\begin{eqnarray}
N(t) & \sim & \bar{N_0} \left[1-\tau\bar{N_0}\left(1+V^2/2 
\right)\right]\nonumber\\
     & &+\bar{N_0}V(1 - 2\tau\bar{N_0})\cos(\omega t+\phi)\nonumber\\
     & & -\,\, \frac{\tau \bar{N_0}^2 V^2}{2} \cos(2\omega t +2\phi)
\label{eq:dft}
\end{eqnarray}
where $\bar{N_0} = Q I_0$.  A DFT at the dither frequency now extracts
an amplitude different than the nominal $\bar{N_0}V$ (and after
division by the nominal mean signal produces an estimate that can
differ significantly from the true visibility amplitude).  However, the
phase comes through the process unaffected, as do the frequencies of
the minima in the visibility amplitudes.  By floating the overall
amplitudes in the data reduction (see below)
we retain the important spatial scale
information contained in the minima.  But the most important conclusion
is that the phases may be assumed to be essentially free of detector
induced biases.

\subsection{Visibility Calibration}\label{subsec:VC}

The degradation of the measured visibilities due to atmospheric and
instrumental effects is measured, as with all interferometers, by
observing calibrator stars with diameters as small as possible to
reduce uncertainties in the visibility estimates for them.  As shown by
\citet{Hummel98}, the NPOI visibility amplitudes sometimes show a
negative correlation with the RMS of the delay line motion which is
related to the seeing. But at other times, instrumental effects which
correlate with time or other systematic effects which correlate with
position on the sky (e.g., hour angle) can dominate the visibility
variations. Therefore, formal photon-noise based visibility errors
usually require the addition (in quadrature) of a calibration error
which is derived from the residual visibility variations of the
calibrator after calibration. For the amplitude calibrations, we
smoothed the calibrator visibilities with a 20\,min Gaussian kernel in
hour angle, and obtained calibration errors ranging from about 4\% at
the red end to 15\% at the blue end of the spectrometers. For the
closure phase, we used the same smoothing technique but applied to the
calibrator phases as a function of time.

During model fitting, however, this standard procedure yielded
inconsistent fits.  One can already see this from the results of
\citet{Ohishi04} (their Fig.\ 6), where amplitudes can be
systematically high or low with the important characteristic that the
deviation is very consistently independent of wavelength. The effect is
exacerbated by the fact that due to the brightness of Altair, the
formal amplitude errors are quite small.  The reason for the
scan-to-scan variations is most likely the same as for the residual
variations of the calibrator after calibration, except that there is no
perfect correlation due to target and calibrator not being at the same
location in the sky. (Past experience has shown that visibilities do
correlate quite well if the calibrator is very near the target.) 
Therefore, we allowed the calibration for each scan and baseline to
float by applying ``achromatic'' calibration factors to improve the fit
between data and model. We will discuss the implications for the model
fitting in Section \ref{modelfitting}. Finally we note that an error in
the calibrator star diameter produces chromatic errors across the NPOI
spectrometer as do uncompensated deadtime corrections, neither of which
can be removed by the calibration factors.

%\input{Modeling}
%\vfill\eject

\section{MODELING}\label{modelfitting}

\subsection{Roche Spheroids}\label{subsec:RS}

The theory for the equilibrium shapes and surface properties of rotating
stars was first presented 80 years ago \citep{vZ24} assuming solid body
rotation and a point source gravitational potential.  This model has
proved quite successful in describing the figures of stars in close
binary systems, where tidal effects to first order produce the same
distortions as rotation \citep{Collins89}.  

We use that model here, but note its limitations.  First, there is no
{\em a priori} reason that stars should rotate as solid bodies, and the
surface layers of the Sun have long been known to rotate
differentially.  However, among the early results of helioseismology
was the discovery that the transition from the outer convection zone to
the inner radiative layers coincided with an abrupt transition to solid
body rotation \citep{SZ92}.  Since early-type stars have radiative
envelopes and relatively small (in radius) convective cores, one might
expect solid body rotation to be a good approximation for the external
layers of early type stars.  

There has been some observational support for this expectation.
\citet{RR04b} have analyzed the rotational profiles of a large number
of A stars looking for evidence of differential rotation following a
solar-type latitudinal dependence.  In the 78 stars for which the
determination could be made, they found 4 objects where pecularities
were seen which might be from differential rotation (or other causes).
However, 95\% of the line shapes were fully consistent with solid body
rotation.

In addition there is the long known consistency between the largest
rotational velocities measured in the early-type stars and the
predicted maximum rotation velocities associated with ``equatorial
breakup'' \citep{Fremat05}.  In recent decades \citep[\eg][]{Tassoul} it
has been demonstrated that rotation laws other than rigid rotation
do not generally impose maximum rotation velocities.

The second limitation, one which we will spend some time on, involves
the exponent in the $T_{eff} - g_{eff}$ relation ($g_{eff}$, the
effective gravity, includes centrifugal terms).  In the original work,
\citet{vZ24} considered the case of a fully radiative envelope,
deriving the well known ``gravity darkening'' relation $T_{eff}
\propto g_{eff}^{0.25}$.  \citet{Lucy67} reconsidered the problem in
fully convective stars, deriving a much reduced gravity dependence,
$T_{eff} \propto g_{eff}^{0.08}$.  Other approximations lead to other
exponents \citep[see][for references]{Reiners03}.  In our nominal
calculations we adopt the original \citet{vZ24} prescription.  As we
shall show, the Altair observations bear significantly on this issue.

Even in the limit of rigid rotation, the \citet{vZ24} theory is only
first order in rotation rate.  Distortions of the interior figure,
allowing some gravitational quadrapole contribution, can be expected as
rotation rates approach breakup.  That is even more likely if there are
significant deviations from solid body rotation, even if confined to
the inner convective regions \citep[\eg][]{Tassoul}.

Probably even more relevant are the effects of radiation pressure,
which is treated simplistically in the theory, and stellar winds.
Significant envelope extension due to radiation can be expected in the
low effective surface gravity regions of rapidly rotating stars.  And
we would certainly expect a dramatic increase in mass loss at the
equator, both due to enhanced convection as gas temperatures decrease
into the solar and sub-solar regime.  We ignore all these effects here,
except to acknowledge the limitations inherent in this theory.

\subsection{Roche Models}\label{subsec:RocheModels}

We have constructed a suite of programs to evaluate the run of specific
intensity across the surface of a Roche spheroid.  The definitions
of the various angles are from \citet{Collins63}.  Otherwise, we follow
the prescription for the surface figure and the notation given by 
\citet{HS68} with one exception: following the discussion by
\citet{HS71} we take the polar radius, $R_p$, as a fixed
parameter. Specifically we do not allow it to be a function of the
fractional rotation.

The modeling requires that we specify six quantities: the ratio of the
angular velocity to that of breakup, $\omega = \Omega/\Omega_B$, the
inclination (or tilt) of the rotational axis, $i$, defined such that
$i=0$ is pole-on, the position angle, $PA$, of the pole on the sky
(measured North through East), the angular diameter of the polar axis,
$\theta_p$, the effective temperature at the pole, $T_p$, and the
surface gravity, or more commonly the logarithm of the surface gravity
(cgs), at the pole, $\log\,g_p$.  From the relations in the cited
references it is then possible to calculate the radius of the star for
a given stellar latitude, $\theta$, and hence the surface gravity
$g_{eff}$ and effective temperature ($T^4(\theta) = T^4_p
(g_{eff}/g_p)$) at that latitude \citep[see][for the definition of
$g_{eff}$]{HS68}.

Note that for stars with accurate parallaxes like Altair, specifying
the polar surface gravity and angular diameter is equivalent to
specifying the mass and linear (polar) radius.  These in turn fix the
breakup angular velocity, $\Omega^2_B = (8/27)GM/R_p^3$, along with the
equatorial and projected velocities (when the inclination is
specified).  Finally, it is useful to recall the relation between the
polar radius and equatorial radius at breakup angular velocity:
$R_{e,B} = 3R_p/2$ \citep{HS68}.  According to this first order theory
the maximum rotational flattening is 2/3.

The model definition is completed by specifying, for each wavelength
and surface point, the specific intensity at the angle of the line of
sight from the local normal.  As noted by the early authors,
plane-parallel model atmospheres are entirely adequate in the context
of Roche models for stars on and near the main sequence.  The only
exception is that these models develop a cusp at the equator at
critical rotation velocity.  However, the cusp does not appear until
fractional rotation velocities of $\omega=0.99$ or larger and, as we
have indicated, the breakdown of the plane-parallel approximation is
only one of several problems with the model in this limit.  A number of
auxillary quantities are also calculated, both integrated over the
spheroid and as seen from that particular direction, in addition to the
predicted complex visibilities.  An example of the former is the
integrated luminosity, calculated as described in
\citet{Collins63}. Examples of the latter include the magnitudes in
various bandpasses.

In practice we need to solve the inverse problem: given a point on the
sky, ($\alpha,\ \delta$), determine whether the point is on the stellar
disk and if so what the corresponding latitude and radius are.  We have
solved this problem explicitly using simple iteration, and first and
second order versions of Newton-Raphson iteration.  The routines,
written in C, are quite flexible, reasonably fast and freely available
(from the first author).

The properties of Roche models for isolated rotating early-type stars
have recently been reviewed by \citet{Dom02}, who use a slightly
different but completely equivalent parameterization.  Those authors
summarize some of the results from their models for massive stars, which
have provided a useful check on our own routines.

\subsection{Implementation of Roche code}

The Roche code consists of a library of functions written in C, with a
main function enabling its use as standalone software and a wrapper
enabling it to be called from within the NPOI standard data reduction
software OYSTER. The Roche spheroid parameters are part of the standard
hierarchical model format of OYSTER and are passed to the Roche code
along with pointers to the extensive tables of linear, logarithmic, and
square-root law monochromatic limb darkening coefficients for a grid of
Kurucz model atmospheres as published by \citet{VH93}. The Roche code,
with the additional input of the $(u,v)$ coordinates, computes the
visibilities for a grid of wavelengths supplied by OYSTER, which are
subsequently integrated over the NPOI bandpasses.

The fitting of the model parameters (except for the gravities) utilizes
the Marquardt-Levenberg algorithm \citep{Press92} implemented in
OYSTER, with the derivatives computed numerically.  In addition to the
visibility and phase measurements the reductions were also constrained
to reproduce the observed V magnitude.  This provided particularly
strong constraints on the polar effective temperature.

%\input{Discussion}
%\vfill\eject

\section{DISCUSSION}\label{Discussion}

\subsection{Model Fitting}

\begin{figure*}
\includegraphics[angle=-90,scale=.65]{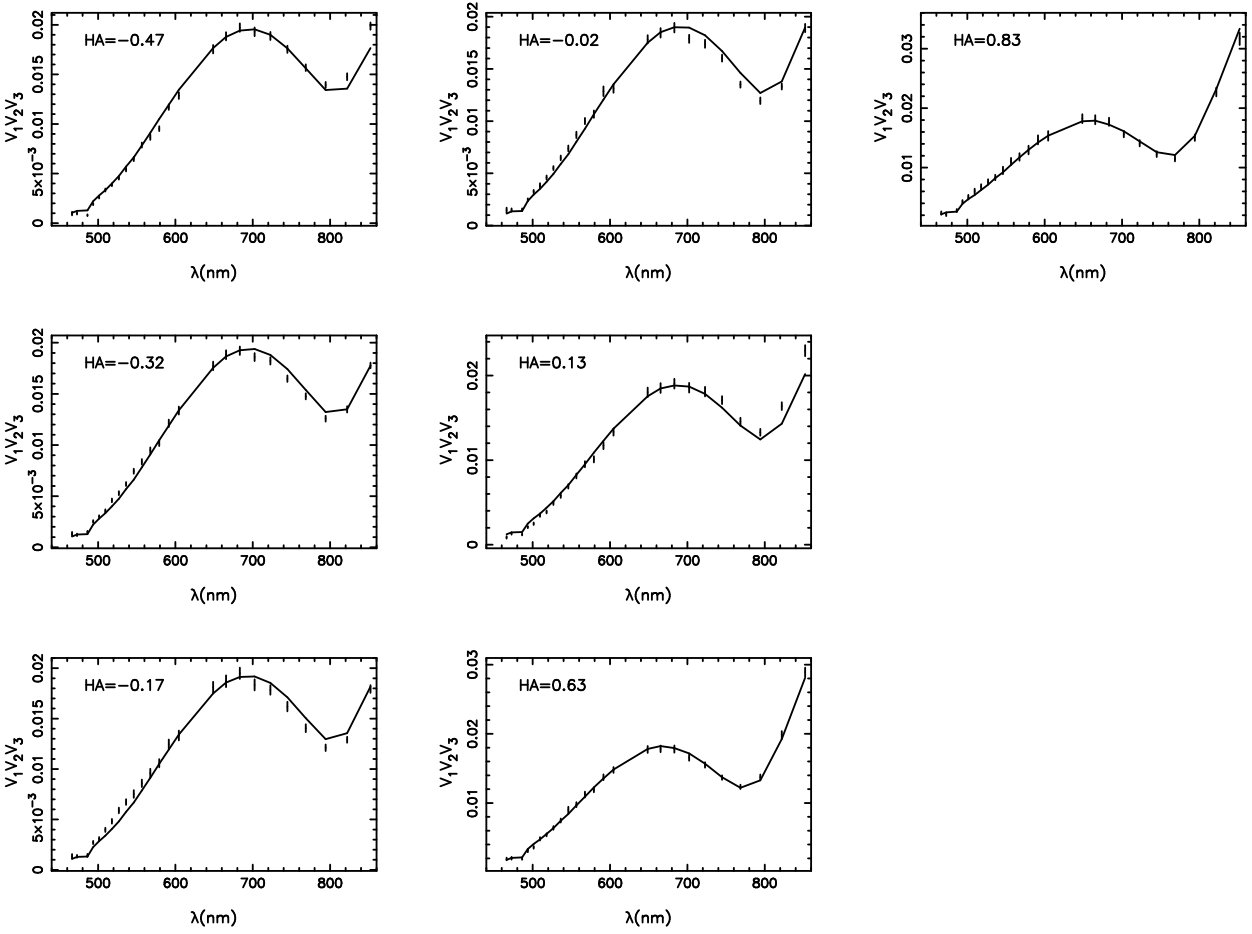}
\caption{Triple amplitudes as fit by best Roche model.  Error bars of 
$\pm\sigma$ are shown. The analytic fits (solid line) include a
constant multiplicative renormalization. \label{Fig:3Amps}}
\end{figure*}

\begin{figure*}
\includegraphics[angle=-90,scale=.65]{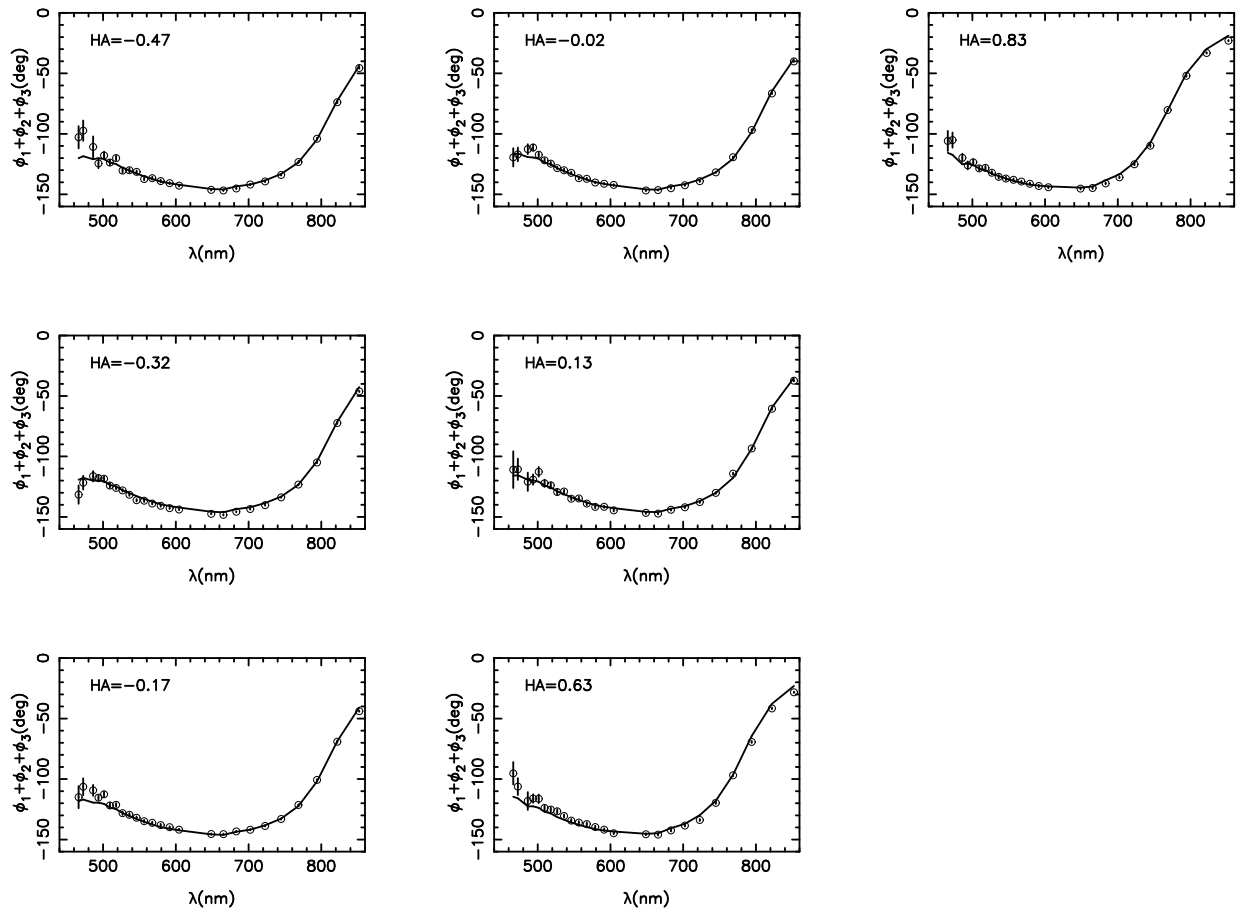}

\caption{Triple phases as fit by the best Roche model.  No
renormalization has been applied. Small circles are used for the
observations since on the red side of the plots the error bars tend to
be smaller than (and fall within) the width of the line showing the
analytic fit.\label{Fig:3Phis}}
\end{figure*}

The analysis proceded with few complications.  In particular, a close
examination of the $\chi^2$ surface indicated no unusual morphology, and
indeed the iterations converged to the same final solution independent
of our starting guess, whether from larger or smaller values of the
parameters. Our final solution, given in column 2 of Table
\ref{Tab:Models}, is based on the triple phases and triple amplitudes
only, the latter with an overall floating multiplier for each scan as
described in \S \ref{subsec:VC}.  The reduced $\chi^2$ for this
solution is still a bit large, which we attribute to some residual
non-gray problems of uncertain origin in the amplitudes, as shown in
Figure \ref{Fig:3Amps}.  The fit to the closure phases, Figure
\ref{Fig:3Phis}, on the other hand is remarkable, showing no trends
with wavelength or hour angle.


The $\chi^2/\nu$ for our best model is well in excess of unity.  This
is the result of both the remaining residuals in the triple amplitudes
evident in Fig. \ref{Fig:3Amps} and the very small formal errors in both
the amplitudes and phases owing to the high signal levels.  If the
large $\chi^2$ had resulted from just the latter, we could have taken
the usual expedient and scaled the errors by a common multiplier
which would have normalized away the excess $\chi^2$.  Estimates of the
uncertainties in the resulting parameters would then have been obtained
through the usual process of varying each parameter until $\chi^2$ had
increased by unity, thereby mapping out out formal errors and the
correlation matrix.

However, this gives very small estimated errors which we feel do not
properly reflect the influence of the (small) remaining biases in the
triple amplitudes.  We have therefore decided to take a conservative
approach to our error estimates.  We have run a separate reduction
using the squared visibilities for the individual baselines in place of
the triple amplitudes.  This solution, which has a substantially larger
reduced $\chi^2$ reflecting the larger residuals in the amplitudes of
the individual baselines, is summarized in column 4 of Table
\ref{Tab:Models}.  Our adopted errors, shown in column 3, are the
difference between these two solutions.

\begin{figure}
\includegraphics[angle=-90,width=3in]{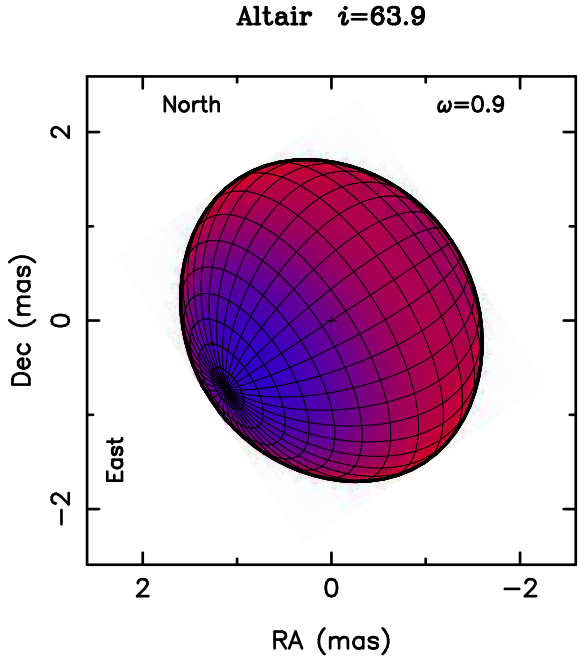}
\caption{A false-color rendering of Altair's visible surface.
Intensity at 500\,nm increases from red to blue.  Except for the
effects of limb-darkening, this is also a map of temperature, which
varies from 8740\,K at the pole to 6890\,K at the equator. 
\label{Fig:Altair}}
\end{figure}

Figure \ref{Fig:Altair} shows how Altair appears projected on the sky.
The intensity distribution at 500\,nm, as would be seen for example by
an interferometer, is color encoded: blue for high intensity, red for
low intensity.  Except for limb-darkening, this is also a temperature
encoding.  The range in intensities, a factor of 18, is about a factor
of 2.5 more than would be expected due to limb-darkening alone in a
non-rotating star of this spectral type.

Included in Table \ref{Tab:Models} is the integrated $B-V$ as calculated
for the models.  Although we force all models to reproduce the observed
V magnitudes, as described above, this constraint does not
automatically mean $B-V$ will be reproduced.  The fact that the color
does agree with the measurements is a significant consistency check on
the models, particularly given that the range of temperatures across
the surface could produce a wide range in $B-V$.  A number of ancillary
parameters derived from the adopted fit are given in Table
\ref{Tab:Altair} and are discussed further in the \S \ref{sec:Conclusions}.

\begin{deluxetable*}{lrrrrcrr}
\tablewidth{0pt}
\tablecaption{Roche Model Fits for Altair and $\zeta$ Aql \label{Tab:Models}}
\tablehead{
&\multicolumn{4}{c}{Altair} & & \multicolumn{2}{c}{$\zeta$ Aql}\\ 
\cline{2-5} \cline{7-8}
&\\[-3mm]
\colhead{Parameter} & \colhead{$V_{123}$ \& $\phi_{\rm cl}\tablenotemark{a}$} 
& \colhead{Errors} 
& \colhead{$V^2_i$ \& $\phi_{\rm cl}\tablenotemark{b}$} & 
\colhead{$\beta = 0.09$} & & 
\colhead{$V^2_i$ \& $\phi_{\rm cl}\tablenotemark{b}$} & \colhead{Errors} 
}
\startdata
$\omega = \Omega / \Omega_c$ & 0.90 & $\pm 0.02$ & 0.88  & 0.978 && 0.990
& $\pm 0.005$\\
$\theta_p$ (mas)             & 2.96 & $    0.04$ & 3.00  &  3.04 && 0.815 
& 0.005\\
$T_p$ (K)     		     & 8740 & $    140 $ & 8600  &  7980 && 
11750 & \nodata\\
$i$ (deg)                    & 63.9 & $     1.7$ & 62.2  &  65.6 && 90 &
+0, -5\\
$PA$ (deg)                   & 123.2& $     2.8$ & 120.4 &  97.4 && 45 &
$\pm 5 $\\
$\chi^2$/DOF                 & 3.8  &    \nodata & 9.5   &  13.4\\
V (obs: 0.77)                & 0.765&    \nodata & 0.765 &  0.76\\
B-V (obs: 0.22)              & 0.215&    \nodata & 0.22  &  0.26\\
$V_{eq}\sin i$ (\kms)        & 245  &    \nodata & 231   &   295\\
\enddata
\tablenotetext{a}{Model fit to the triple amplitude and closure phase data.}
\tablenotetext{b}{Models fit to the closure phase data and the squared
visibilities of the three baselines.}
\end{deluxetable*}

\subsection{Comparison with Previous Results}

As noted, Altair has been the subject of a number of interferometric
measurements over the years, most notably by the PTI array
\citep{vBelle01}.  The diameter determinations are not simply
intercomparable, even if we consider only the projected major and minor
axes of our model, since the PTI observations were fit to the physical
dimensions of a Roche model, which included limb-darkening but not
gravity darkening.  Still, our determination of an equatorial radius of
1.988\,\Rsun\ seems in reasonable accord with their quoted 1.88\,\Rsun.

More problematic are the reported position angles of the rotational
pole. First, we note that in a preliminary report of this work
\citep{P-SPIE04} the pole is off by $180^{\circ}$ due to a sign error.
More complicated is the disagreement between the position angle from
the PTI measurements, $-25^{\circ}\pm 9$ compared to our $123\ddeg2\pm
2\ddeg8$.  Communication with van Belle (2003, private communication)
indicated that the $(u,v)$ coordintates of each baseline were
inadvertantly exchanged in their analysis.  Fitting a simple uniform
ellipse model to the PTI data, corrected for the component swap, yields
$PA=122\ddeg2$, in excellant aggrement with our result.

Recently \citet{RR04b} have reported the determination of Altair's
equatorial rotational velocity, $v_{eq} \leq 245$\,\kms.  They analyzed
the star's rotational broadening profile to determine the first two
zeros of its Fourier transform, the ratio of which has been shown by
\citet{Reiners03} to depend on the equatorial velocity rather than the
usual $v \sin i$.  This is a new approach to measuring total velocities
in stars, and it is difficult to know how much weight it should be
given.  One notable aspect of that analysis was the adoption of an
exponent for the gravity darkening law ($\beta \sim 0.09)$ that was
about $1/3$ that of the \citet{vZ24} value.  We discuss this aspect of
calculating rotationally distorted stars next.

\subsection{Gravity Darkening}
While we have several lines of reasoning, described in \S
\ref{subsec:RS}, that lead us to believe that solid body rotation is
valid for these models, the situation is not so clear with regard to
the exponent on the gravity darkening law, $T_{eff} \propto
g_{eff}^{\beta}$.  The classical work of \citet{vZ24} would seem to
apply to an A star, even one somewhat evolved, yielding $\beta = 0.25$.
However, convection does occur in late A stars, and particularly in the
photosphere where it competes with radiation in carrying the flux.  It
is then important to note that \citet{Lucy67} has shown that for small
distortions the appropriate coefficient in fully convective envelopes
is closer to $\beta \simeq 0.08$.

This leaves matters in a somewhat uncertain state.  On the one hand,
there are stars in the transition region between having fully
convective and fully radiative envelopes, Altair arguably one of them,
and there is no obvious guidance in choosing an appropriate value for
this parameter.  On the other hand, even where the envelopes are
unambigiously in either one of those states or the other, the classical
results apply under rather different circumstances, which we next
discuss.

Both the \citet{vZ24} and \citet{Lucy67} results treat rotation as a
perturbation.  However, in the radiative case where uniform rotation is
adequate (and issues like mass loss, etc.\ can be ignored), the quantity
treated as a perturbation is the size of the quadrapole moment of the
gravitational field.  Since stars are centrally condensed, even for
velocities approaching critical the distortions in the core are modest,
and one can expect that the analysis given by \citet{vZ24} will be
reasonably accurate.  This has been found to be true in practice
\citep{Sackmann70}.

For the convective case, the gravity darkening exponent is obtained by
analyzing the adiabats found in the envelopes of representative stars.
\citet{Lucy67} quite explicitly points out that the derivation is valid
only for small changes in the {\em effective gravity}.  Of course, the
effective gravity changes by orders of magnitude as rotation approaches
critical, and it is not clear whether the exponent derived by Lucy can
be used to describe gravity darkening for anything but the most modest
rotation.  Again, this is in contrast to the modest contribution of an
induced gravitational quadrapole, even for stars rotating at breakup.

Even so, in a recent series of papers, \citet[and references
therein]{Claret04} has attempted to deal with the issue of a smooth
interpolation between these two extreme cases.  He has noted that as
stars evolve off the main sequence and toward the red giant branch,
their interior structures trace out approximately straight line loci in
a $(\log T_{eff}, \log g)$ diagram.  On the main sequence for massive
(mostly radiative) stars the slope of this line is about 0.25, and for
intermediate mass stars ($\sim 1$\,\Msun, mostly convective) the slope
is about 0.06, the two values being remarkably close to the radiative
and convective exponents cited above.  Working with the interiors
codes, \citet{Claret04} is able to evaluate this exponent at each point
in the evolutionary paths of models covering $40 \geq \Msun \geq 0.08$,
offering the results as appropriate exponents to use in rotating stars
and stars in close binary systems over their entire evolutionary
lifetimes.

This is a constructive suggestion for the thorny problem of choosing an
appropriate gravity darkening law.  However, we are not fully convinced
of the leap of going from deriving a quantity based on evolutionary
changes to using it to describe the effects of rotational distortion.
One might make the case for small amounts of rotation or if it could be
shown that rotational distortions and evolutionary effects were close
to being homologous transformations from one to the other.  But
rotational distortions are not homologous to evolutionary changes and
it is not at all clear how well these ``interpolations'' work.

The observations reported here bear somewhat on this problem.  We have
tried converging our Roche models using the value $\beta = 0.09$
adopted by \citet{RR04b}.  The results are shown in column 5 of Table
\ref{Tab:Models}.  The $V^2$ data set was used so the comparison is
with column 4.  To achieve the degree of asymmetry found in the triple
phase data with this low exponent value, the rotation parameter is
forced to near critical rotation, $\omega \sim 0.978$. In turn the
predicted projected rotational velocity, $v \sin i \sim 295$\,\kms\
conflicts with the observed value, and the predicted color is
significantly redder than observed.  Further, the reduced $\chi^2$ is
significantly worse for this fit.

We feel it is premature to use these observations to derive a ``best''
value of the gravity darkening parameter until the remaining visibility
residuals are better understood.  Since it is the phase measurements,
which we trust, which are sensitive to asymmetries in brightness across
the disk and not the visibility amplitudes, it does appear that the
\citet{vZ24} value for that parameter is superior to the value adopted
by \cite{RR04b}.

Moreover, as we will describe in \S \ref{subsec:Status}, our results
indicate that Altair has hardly evolved from the zero age main
sequence.  The \citet{Claret98, Claret04} tables give values for $\beta$ in
essential agreement with the \citet{vZ24} result during this phase of
evolution.  Thus, Altair does not provide a test of those tables, but
does highlight a difficulty in applying the technique proposed by
\citet{RR04b} for finding total velocities, namely the difficulty of
obtaining {\it a priori} reliable estimates for $\beta$.

\subsection{H$\beta$}

\begin{figure}
\includegraphics[angle=-90,width=3in]{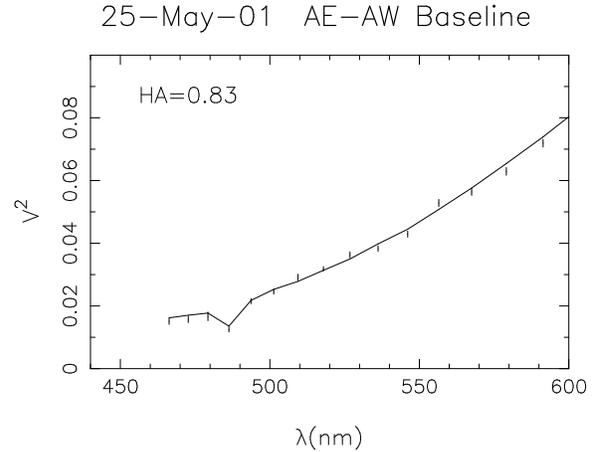}
\caption{One of the observed V$^2$'s for the AE--AW baseline. The effect
of the strong H$\beta$ feature in the $\lambda$\,468.3 channel is
clearly evident and well matched by the model. \label{Fig:Hbeta}}
\end{figure}

In Figure \ref{Fig:Hbeta} we show the blue squared visibilities of the
AE-AW baseline for the $HA=0\dhrs83$ observation.  The notable feature
at 486\,nm is H$\beta$; the agreement with the calculations shown in
this scan, and the others not shown, is striking. This feature is
nearly centered in the 486.3\,nm channel.  In contrast, H$\alpha$,
which has a smaller equivalent width and is split between channels at
$\lambda$\,665.4 and $\lambda$\,648.7, is much less noticable.  The
main reason for the feature being reflected in the amplitudes is the
reduction in the limb-darkening coefficient, the star appears to be
more like a uniform disk at this wavelength, and thus the visibility is
reduced.  This close agreement is a nice confirmation of the details of
the model fits.

%\input{Conclusions}
%\vfill\eject

\section{Conclusions}\label{sec:Conclusions}

\subsection{Imaging Altair}
The primary result of these observations is, we believe, the first
detection of asymmetric surface intensities on the surface of a star
induced by rotation.  Although we have imposed a model on the data and
fit the model parameters, the simple conclusion, first reported by
\citet{Ohishi04}, is that the surface of Altair displays an extremely
asymmetric intensity distribution and that the asymmetry is consistent
with that expected from the known high rotation and with the previously
reported oblateness \citep{vBelle01}.  We have, in effect, imaged the
surface of an A star.

In Table \ref{Tab:Altair} we summarize various physical parameters for
the adopted model --- column 2 of Table \ref{Tab:Models}.  Most
quantities should be self-explanatory. Subscript ``B'' refers to the
model if it were rotating at breakup. The angular diameter $\theta_{min}$
is for the projected minor axis while $\theta_{Max}$ is for the projected
major axis, \ie\ the angular diameter of the equator.

\subsection{The Status of Altair}\label{subsec:Status}

Knowing now the rotational state of Altair, we can better answer
fundamental questions such as its evolutionary status.  Over the years
Altair has been classified as A7 IV-V \citep[\eg][]{JohnsonMorgan53},
the luminosity class usually indicating an object slightly past the end
of its main sequence evolution while the spectral type is that of a
star having an effective temperature in the vicinity of 7800K
\citep[\eg][]{ErspamerNorth03,vBelle01}.  In the context of analyzing
Altair's pulsations, it is important to know in detail its evolutionary
state, mainly the extent of its core, and to remove any biases that
might be introduced by rotation.

Fortunately, in the context of rigid rotation, this is not so
difficult.  Early results \citep[\eg][]{Sackmann70} showed that two
quantities were relatively insensitive to the effects of rotation:
polar radius and total luminosity.  The Roche model fits give polar
radius directly while it is a straightforward matter to calculate the
total luminosity \citep[\eg][]{Collins63}.  

These two quantities are not perfectly conserved.  In the range of
interest, $3\,\Msun \geq M \geq 1.4\,\Msun$, both quantities decrease
with increasing rotation, approximately in proportion to $\omega^2$,
reaching a maximum correction of about 6\% in luminosity and 1.5\% in
radius \citep{Sackmann70}. For stars in the neighborhood of
$1.8\,\Msun$, rotating with $\omega^2\sim 0.8$ we find from the
\citet{Sackmann70} calculations that the non-rotating star would be 4\%
more luminous and 1\% larger than our deduced polar radius.

To estimate the parameters of the appropriate non-rotating star we have
used the evolutionary tables by the Geneva group \citep{Schaller92}.
These models were calculated with modest convective overshoot
($0.2H_p$).  We have used the grid for a composition of $X=0.68$,
$Y=0.30$, $Z=0.02$. 

\begin{deluxetable}{lcrr}
\tablewidth{0pt}
\tablecaption{Altair Physical Parameters\tablenotemark{a}\label{Tab:Altair}}
\tablehead{
\colhead{Quantity} & \colhead{Unit} & \colhead{Value} & \colhead{Error}}
\startdata
\cutinhead{Rotating Model Parameters}\\
V$_{eq}$ & \kms & 273 & 13\\
V$_{eq,B}\tablenotemark{b}$ & \kms & 374 & 3\\
$\Omega$ & day$^{-1}$ & 2.71 & 0.11\\
$\Omega_B\tablenotemark{b}$ & day$^{-1}$ & 3.01 & 0.06\\
T$_p$ & K & 8740 & 140\\
T$_{eq}$ & K & 6890 & 60\\
R$_p$ & \Rsun & 1.636 & 0.022\\
R$_{eq}$ & \Rsun & 1.988 & 0.009\\
$\theta_{min}$ & mas & 3.056 & 0.047\\
$\theta_{Max}$ & mas & 3.598 & 0.017\\
$\log L$ & \Lsun & 1.027 & 0.011\\
$\log g_p$ & cgs & 4.266 & 0.012\\
$\log g_{eq}$ & cgs & 3.851 & 0.035\\
\cutinhead{Non-rotating Parameters\tablenotemark{c}}\\
M & \Msun & 1.791 & 0.018\\
R & \Rsun & 1.652 & 0.022\\
$\log g$ & cgs & 4.256 & 0.002\\
$T_{eff}$ & K & 8200 & 98\\
$X_c$ & -- & 0.607 & 0.019\\
\enddata
\tablenotetext{a}{Uncertainties due to the parallax have not been
included in the errors.}
\tablenotetext{b}{Rotating at breakup but with the same mass and polar
radius.}
\tablenotetext{c}{The parameters of a non-rotating star from
the Geneva grid \citep{Schaller92} which would reproduce the
(corrected) luminosity and polar radius -- see text.}
\end{deluxetable}

The quantities given in Table \ref{Tab:Altair} as ``Non-rotating'' are
those estimated from the \citet{Schaller92} models and are quite
striking.  The last entry in Table \ref{Tab:Altair} is the mass
fraction of hydrogen remaining in the core.  This is to be compared to
a starting value of $X_c=0.68$.  Altair is almost on the ZAMS.

\subsection{Chromospheric indicators}
As mentioned in \S \ref{Introduction}, Altair is one of two A7 objects,
the other being $\alpha$\,Cep, in which certain ultraviolet emission
lines, taken as indicators of chromospheric temperature inversions, are
seen.  No objects of earlier spectral type are known to show these
features, and it is usually argued that these therefore represent the
hottest photospheres where convection is still capable of creating such
temperature profiles.  The model for Altair adopted here calls that
conclusion into question.  As shown in Table \ref{Tab:Altair} Altair
has a broad swath of 6900\,K gas at its equator, which is the likely
source of the strong convection.  We also note the recent announcement
\citep{vBelle05} that substantial oblateness and gravity darkening have
been found in $\alpha$\,Cep, suggesting significant amounts of cool,
convective gas in that object as well.

\subsection{A $\delta$\,Scuti star}
As we also indicated in \S \ref{Introduction}, rather than just being
another rotating A star, Altair may prove to be a valuable laboratory
for examining the internal rotation state of a star with a
predominantly radiative envelope.  \citet{Buzasi05} have announced the
discovery of $\delta$\,Scuti pulsations in Altair and have identified
several of the periods with frequencies mostly in the range 15 -- 18
day$^{-1}$. Interestingly, two of the frequencies reported by
\citet{Buzasi05}, 3.526\,day$^{-1}$ and 2.57\,day$^{-1}$ were
significantly lower, the latter being quite close to the rotational
frequency we have derived, \eg Table \ref{Tab:Altair}.

This was immediately recognized as providing a potential probe of the
interior structure and particularly the rotation law, and an attempt
has been made to identify and model the modes \citep{Suarez05}.
Unfortunately, not knowing the rotational state of the star and making
the assumption that equator-on was the most likely orientation,
\citet{Suarez05} adopted a total rotation significantly lower than now
seems likely.  Other effects of this choice included identifying the
evolutionary state as being substantially more advanced than we believe
is the case.  As is clear from their results, rotational velocities
above the 180--240\,\kms\ range they investigated lead to large changes
in the oscillation modes, making mode identification difficult.  It
will not be an easy task to tap the information being provided by
Altair.

\acknowledgments

The referee, Gerard van Belle, asked a number of penetrating questions
that significantly improved this work.  Notably, he wondered about the
effects of rotation on our calibrators, leading to the discovery that
we could measure those effects in $\zeta$ Aql.  The NPOI facility is a
collaboration between the Naval Research Laboratory and the US Naval
Observatory in association with Lowell Observatory, and was funded by
the Office of Naval Research and the Oceanographer of the Navy.  This
research has made use of the SIMBAD literature database, operated at
CDS, Strasbourg, France, and of NASA's Astrophysics Data System.

%Facilities: \facility{NPOI}.

%\input{References}

%\vfill\eject
%\input{Appendix}
%\vfill\eject
%\input{Figures}
%\vfill\eject
%\input{Tables}
\end{document}